\documentclass[final,5p,times,twocolumn]{elsarticle}
\usepackage{amssymb,amsmath,amsthm}
\usepackage{mathrsfs}
\usepackage{mathtools}
\usepackage{graphicx,epsfig,url}
\usepackage[caption=false,font=footnotesize]{subfig}
\usepackage{microtype}
\usepackage{tabularx}
\usepackage{xcolor}

\usepackage{enumitem}

\newcommand{\norm}[1]{\lVert#1\rVert}

\def\n{\boldsymbol{n}}

\def\v{\boldsymbol{v}}
\def\u{\boldsymbol{u}}

\def\W{\mathbf{W}}

\def\B{\mathbf{B}}
\def\C{\mathbf{C}}

\def\I{\mathbf{I}}
\def\P{\mathbf{P}}

\def\w{\boldsymbol{w}}
\def\x{\boldsymbol{x}}
\def\z{\boldsymbol{z}}

\def\y{\boldsymbol{y}}

\newtheorem{theorem}{Theorem}
\newtheorem{definition}{Definition}

\allowdisplaybreaks

\begin{document}

\begin{frontmatter}
\title{Compressive Sensing of ECG Signals using Plug-and-Play Regularization\tnoteref{label1}}
\tnotetext[label1]{K.~N.~Chaudhury was partially supported by Core Research Grant CRG/2020/000527 and SERB-STAR Award STR/2021/000011 from the Department of Science and Technology, Government of India.}

\author{Unni V. S.\corref{corresponding}}
\cortext[corresponding]{Corresponding author.}
\ead{unniv@iisc.ac.in}
\author{Ruturaj G. Gavaskar}
\ead{ruturajg@iisc.ac.in}
\author{Kunal N. Chaudhury}
\ead{kunal@iisc.ac.in}

\affiliation{organization={Department of Electrical Engineering, Indian Institute of Science},
            city={Bangalore},
            postcode={560012}, 
            country={India}}

\begin{abstract}
Compressive Sensing (CS) has recently attracted attention for ECG data compression. In CS, an ECG signal is projected onto a small set of random vectors. Recovering the original signal from such compressed measurements remains a challenging problem. Traditional recovery methods are based on solving a regularized minimization problem, where a sparsity-promoting prior is used. In this paper, we propose an alternative iterative recovery algorithm based on the Plug-and-Play (PnP) method, which has recently become popular for imaging problems. In PnP, a powerful denoiser is used to implicitly perform regularization, instead of using hand-crafted regularizers; this has been found to be more successful than traditional methods. In this work, we use a PnP version of the Proximal Gradient Descent (PGD) algorithm for ECG recovery. To ensure mathematical convergence of the PnP algorithm, the signal denoiser in question needs to satisfy some technical conditions. We use a high-quality ECG signal denoiser fulfilling this condition by learning a Bayesian prior for small-sized signal patches. This guarantees that the proposed algorithm converges to a fixed point irrespective of the initialization. Importantly, through extensive experiments, we show that the reconstruction quality of the proposed method is superior to that of state-of-the-art methods.
\end{abstract}

\begin{keyword}
ECG signals, compressive sensing, proximal gradient descent, plug-and-play regularization, GMM denoiser.
\end{keyword}

\end{frontmatter}

\section{Introduction}

Electrocardiogram (ECG) is widely used for diagnosis and monitoring cardiac conditions such as hypertension \cite{ahmad2012electrocardiogram}, heart failure \cite{pecchia2010discrimination}, arrhythmia \cite{owis2002study} etc.
Essentially, an ECG signal is a representation of the electric activity in the heart over time.
Extensive use of ECG in healthcare fosters a need for sophisticated signal processing approaches to efficiently compress, analyze, store and transmit ECG signals. 
For example, in wearable devices, there is a need to reduce energy consumption due to data transmission and to increase memory usage efficiency. 
Recently, several efforts have been made to develop wireless  ECG sensors for continuous health monitoring, for which it is desirable to have devices with low power consumption or low complexity.
However, continuous wireless transmission of long-term biomedical data consumes a significant amount of energy.
Thus, compression of ECG signals would be helpful to achieve energy efficiency. 

Compressive Sensing (CS) is a possible solution for signal compression.
It employs random linear projections that aim to preserve the structure of the signal.
The signal can be reconstructed from its projections using nonlinear recovery methods.
In fact, several works have explored the application of CS to biomedical signal processing, including ECG \cite{mamaghanian2011compressed, zhang2012compressed, pant2013compressive}, EMG \cite{salman2011compressive, dixon2012compressed}, EEG \cite{aviyente2007compressed} signals and MRI images \cite{lustig2007sparse}.
\subsection{Compressive Sensing of ECG Signals}

The data acquisition model in CS is given by \cite{zhang2012compressed,pant2013compressive}
\begin{equation}
\y = \boldsymbol{\Phi} \x + \n
\label{eq:model}
\end{equation}
where $\x \in \mathbb{R}^{N}$ is the original ECG signal having length $N$, $\boldsymbol{\Phi} \in \mathbb{R}^{M \times N}$ is a compression matrix with $M \ll N$, and $\n$ denotes the noise in the acquisition system.
Typically, $\n$ is assumed to be white Gaussian noise with mean $0$, whereas $\boldsymbol{\Phi}$ is taken to be a random Gaussian or binary matrix \cite{mamaghanian2011compressed, zhang2012compressed, pant2013compressive}.

The original signal $\x$ can be approximately recovered by solving the regularized inversion problem
\begin{equation}
\label{unconstrained}
\underset{\x \in \mathbb{R}^N}{\text{min}} \ \ f(\x) +  \lambda g(\x),
\end{equation} 
where the term $f(\x) = \frac{1}{2}\Vert \y - \boldsymbol{\Phi} \x \Vert^2$ forces consistency of the recovered signal w.r.t. the measurements, whereas $g(\x)$ (known as the regularizer) acts as a penalty function that forces the recovered signal to have some desirable properties such as smoothness.
Here, $\lambda$ is a positive scalar used to control the amount of regularization and $\lVert \cdot \rVert$ denotes the $\ell_2$ norm.

A good regularizer $g(\x)$ is necessary since recovering $\x$ from $\y$ is an ill-posed problem (as $M \ll N$).
Several regularizers have been explored for the ECG compressive recovery task, such as the weighted $\ell_1$ norm \cite{polania2014weighted}, various other $\ell_p$ norms \cite{pant2014new}, total variation (TV) \cite{liu2013multi}, and second-order sparsity-promoting functions \cite{pant2013compressive}.
Moreover, efficient recovery algorithms have been derived by exploiting the temporal correlation between successive samples; examples include $\ell_{p}^{1d}$-regularized least-squares \cite{pant2013reconstruction}, Block Sparse Bayesian Learning Bound-Optimization (BSBL-BO) \cite{zhang2011sparse}, and Block Sparse Bayesian Learning with Expectation Maximization (BSBL-EM) \cite{zhang2013extension}.
The latter two are considered to be state-of-the-art.

\subsection{Classical Regularization}

It is well-known that the $\ell_1$ norm promotes sparse solutions; moreover, natural signals such as ECG are known to be approximately sparse in suitably chosen domains \cite{polania2014weighted,liu2013multi}.
Therefore, sparsity-promoting regularizers based on the $\ell_1$ norm in the wavelet and gradient domains (TV) have traditionally been used for ECG reconstruction.
The downside is that the $\ell_1$ norm is not differentiable.
Hence, the objective function in \eqref{unconstrained} as a whole is non-differentiable, even though $f(\x)$ is differentiable.
A good choice of iterative numerical solvers to solve such problems is the class of \textit{proximal algorithms} \cite{Beck2017_optimization}, such as ADMM and Proximal Gradient Descent (PGD).
A proximal algorithm generally consists of smaller subproblems which individually involve only one of the two functions.
Consequently, it is possible to take advantage of the differentiability of $f(\x)$.
In this paper, we focus on PGD since it is a particularly simple proximal algorithm.
PGD is sometimes known as the Iterative Shrinkage-Thresholding Algorithm (ISTA) \cite{BT2009}.
Starting from an initial point $\x_0 \in \mathbb{R}^N$, PGD creates a sequence of points $\x_1,\x_2,\ldots$ recursively using the rule
\begin{equation}
\label{ista}
\x_{k+1} = \mathrm{prox}_{\gamma \lambda g} \big( \x_k-\gamma\nabla f(\x_k) \big),
\end{equation}
where $\gamma > 0$ is a fixed parameter (known as the step size) and $\mathrm{prox}_{\gamma \lambda g}(\cdot)$ is a function known as the \textit{proximal operator} of $g$:
\begin{equation}
\label{prox_map}
\mathrm{prox}_{\gamma \lambda g}(\z) = \mathop{\mathrm{argmin}}_{\w \in \mathbb{R}^N} \left[ \frac{1}{2}\Vert \w - \z \Vert^2 + \gamma \lambda g(\w) \right].
\end{equation}
Note that if we put $\boldsymbol{\Phi} = \I$ in \eqref{unconstrained}, then \eqref{unconstrained} reduces to \eqref{prox_map}.
Thus, the proximal operator can be interpreted effectively as a Gaussian denoising operator.
For regularizers such as the $\ell_0$ and $\ell_1$ norms, the proximal operator has a closed-form formula \cite{Beck2017_optimization}.
Hence, the PGD algorithm is easy to implement.
As discussed in Section \ref{sec:pnp}, the algorithm is guaranteed to converge under some mild conditions to a minimum of $f(\x) + \lambda g(\x)$.
Note that every PGD step can be seen as the composition of two operations: the first (computing $\x_k - \gamma\nabla f(\x_k)$) is effectively one step of the classical gradient descent algorithm and depends only on the function $f$, while the second (computing the proximal operator) depends only on $g$.
This is why the algorithm is named as Proximal Gradient Descent.

\subsection{Plug-and-Play Regularization}

Plug-and-play (PnP) regularization is a novel regularization technique developed in the image processing community \cite{venkatakrishnan2013plug,kamilov2017plug}.
The main step in PnP is to replace the proximal operator by a powerful signal denoiser.
As discussed in Section \ref{sec:pnp}, this is due to the similarity of the proximal operator with a denoising operation.
In the context of PGD, the function $\mathrm{prox}_{\gamma \lambda g}(\cdot)$ is replaced by a signal denoiser $D(\cdot)$, so that the $k$-th step now becomes $\x_{k+1} = D\big( \x_k-\gamma\nabla f(\x_k) \big)$.
This algorithm is known as PnP-PGD.
Essentially, this amounts to taking one step of gradient descent, followed by denoising.
Note that we no longer need to choose a regularizer $g(\x)$ since $g$ does not appear in the modified algorithm; the regularization is performed implicitly by the denoiser.

PnP has yielded state-of-the-art results in many imaging problems.
However, since there is no regularizer $g$ involved, the aforementioned convergence to a minimum of $f(\x) + \lambda g(\x)$ does not apply.

\subsection{Contribution}

The contributions of this work are as follows.
\begin{enumerate}[label=\arabic*., labelwidth=!, labelindent=0pt]
\item We introduce the PnP framework for reconstructing ECG signals from CS measurements.
To the best of our knowledge, PnP has never been used for this application.
Through extensive experiments, we show that the proposed method, based on PnP-PGD, outperforms the current state-of-the-art CS recovery methods for ECG signals.
\item Even though convergence of PGD to a minimum of $f + \lambda g$ is not applicable to PnP-PGD, we show that a different form of convergence, known as \textit{fixed-point convergence}, can be guaranteed if the ECG denoiser $D(\cdot)$ satisfies a condition known as \textit{contractivity}.
Thus, the challenge lies in designing a contractive ECG denoiser.
\item We derive a high-quality contractive ECG denoiser $D(\cdot)$ by modeling small patches as random vectors following a Gaussian Mixture Model (GMM).
We experimentally show that the denoising performance of the GMM denoiser is comparable or better than existing state-of-the-art ECG denoisers.
\end{enumerate}

The rest of this paper is organized as follows.
In Section \ref{sec:pnp}, we give an overview of the PnP-PGD algorithm and explain the motivation behind its development.
We derive the GMM denoiser in Section \ref{sec:denoiser} and compare its denoising quality with existing ECG signal denoisers.
In Section \ref{sec:conv}, we discuss how the this denoiser can be used in a way that guarantees convergence of PnP-PGD for CS recovery.
Numerical experiments on CS recovery of ECG signals are shown in Section \ref{sec:experiments}, and we conclude the paper in Section \ref{sec:conclusion}.
Some of the mathematical proofs are given in the Appendix.

\section{Plug-and-Play PGD}
\label{sec:pnp}

We first state a standard convergence result for PGD and then move on to discuss some convergence-related aspects of PnP.
What makes PGD a simple yet powerful algorithm is its guarantee of convergence to a minimum of $f(\x) + \lambda g(\x)$.
In the following theorem (and the rest of the paper), we denote the largest singular value of a matrix by $\sigma_{\mathrm{max}}(\cdot)$.
\begin{theorem}[\cite{BT2009}]
\label{ista-conv}
Consider the PGD algorithm for minimizing the function $f(\x) + \lambda g(\x)$, where $f(\x) = \frac{1}{2}\Vert \y - \boldsymbol{\Phi} \x \Vert^2$
Suppose $g$ is continuous and convex, and that $0 < \gamma < 2/\sigma_{\mathrm{max}}(\boldsymbol{\Phi}^\top \boldsymbol{\Phi})$.
Then the sequence $f(\x_k) + \gamma g(\x_k)$ converges to the minimum of $f(\x) + \lambda g(\x)$ as $k \to \infty$.
\end{theorem}

We now turn our attention to the PnP framework.
Consider the definition of $\mathrm{prox}_{\gamma \lambda g}(\z)$ in \eqref{prox_map}.
Note that the minimization problem in \eqref{prox_map} is similar to \eqref{unconstrained} if we put $\boldsymbol{\Phi} = \mathbf{I}$, the identity matrix.
Hence, $\mathrm{prox}_{\gamma \lambda g}(\z)$ is essentially a regularized inverse corresponding to the additive noise model $\z = \w + \n$, where $\n$ is Gaussian noise.
Thus, the proximal operator is simply an additive Gaussian denoising operator.

It is well-known in the image processing community that specially designed denoisers such as nonlocal means (NLM) \cite{Buades2005} and BM3D \cite{Dabov2007} are superior to traditional denoisers based on regularization, e.g. $\ell_1$ or TV-regularized denoising.
Motivated by this observation, the work in \cite{venkatakrishnan2013plug} explored how the performance of a proximal algorithm for image recovery problems is affected if we replace $\mathrm{prox}_{\gamma \lambda g}(\cdot)$ by some arbitrary Gaussian denoiser $D(\cdot)$, such as NLM or BM3D.
This scheme was named as plug-and-play, since the denoiser $D$ serves as a pluggable module that replaces the proximal operator in an already existing numerical solver.
In the original work \cite{venkatakrishnan2013plug}, the PnP scheme was explored for a different proximal algorithm -- ADMM -- but it was adapted to PGD in \cite{kamilov2017plug}.
The PnP-PGD algorithm is thus recursively defined by
\begin{equation}
\label{pnp-ista-general}
\x_{k+1} = D \big( \x_k-\gamma\nabla f(\x_k) \big).
\end{equation}
Note that the same denoiser $D$ can be utilized for several kinds of image recovery problems using the PnP framework, since only the function $f$ changes from problem to problem.
For this reason, in the past few years, PnP has gained a lot of interest in the imaging community.
However, the use of PnP for recovering one-dimensional signals such as ECG signals has remained an unexplored territory.

An immediate question that arises from the PnP scheme is as follows:
Does the sequence $(\x_k)$ converge to some $\x^\ast$?
And if so, is $\x^\ast$ optimal in some sense?
The latter question can be resolved if $D(\cdot)$ is expressible as the proximal operator of some function $g$.
In general, however, an arbitrary $D$ cannot be expressed in this way.
As a result, the PnP-PGD algorithm cannot be interpreted as minimizing an objective function of the form $f + \lambda g$, and the convergence result in Theorem \ref{ista-conv} is not generally applicable.
Therefore, we are left with trying to determine whether at least the sequence $(\x_k)$ converges.
It turns out that such a guarantee can indeed be given under a technical condition on $D$.
\begin{definition}
\label{contractive-map}
The denoiser $D : \mathbb{R}^N \to \mathbb{R}^N$ is said to be contractive if there exists $\delta \in [0,1)$ such that for all points $\z_1,\z_2 \in \mathbb{R}^N$,
\begin{equation*}
\lVert D(\z_1) - D(\z_2) \rVert \leqslant \delta \lVert \z_1 - \z_2 \rVert
\end{equation*}
\end{definition}
We now state a theorem that guarantees the convergence of PnP-PGD using a contractive denoiser.
\begin{theorem}
\label{pnp-ista-conv}
Consider the PnP-PGD algorithm, $\x_{k+1} = D \big( \x_k - \gamma \nabla f(\x_k) \big)$, where $f(\x) = (1/2) \lVert \boldsymbol{\Phi} \x - \y \rVert^2$.
Suppose $0 < \gamma \leqslant 2/\sigma_{\mathrm{max}}(\boldsymbol{\Phi}^\top \boldsymbol{\Phi})$.
Moreover, suppose the denoiser $D$ is contractive.
Then, as $k \to \infty$, the sequence $\x_1,\x_2,\ldots$ converges linearly (at an exponential rate) to a unique fixed point $\x^\ast$ that does not depend on the initialization $\x_0$.
\end{theorem}

While Theorem \ref{pnp-ista-conv} is proved in the Appendix, we mention here that the proof uses the Banach Fixed Point Theorem \cite[Th. 9.23]{W1976}, from which the linear rate of convergence follows.

Note the difference between the types of convergence addressed in Theorems \ref{ista-conv} and \ref{pnp-ista-conv}:
Theorem \ref{ista-conv} claims the convergence of the sequence of objective function values $f(\x_k) + \lambda g(\x_k)$, whereas Theorem \ref{pnp-ista-conv} is concerned with the sequence of variables $\x_k$.
Theorem \ref{pnp-ista-conv} essentially claims that the PnP-PGD algorithm eventually stabilizes, in the sense that two consecutive iterates $\x_k$ and $\x_{k+1}$ are close to each other.
This property is known as fixed-point convergence \cite{NGC2021}, and is desirable for any recovery algorithm.
Thus, by Theorem \ref{pnp-ista-conv}, it is sufficient for the denoiser $D$ to be contractive, in order to have fixed-point convergence.

It is useful to compare our result with a similar result in a recent work \cite{ryu2019plug}.
In \cite{ryu2019plug}, the authors proved fixed-point convergence of PnP-PGD under a different set of assumptions than ours.
The convergence result in \cite{ryu2019plug} is applicable to the case where the loss function $f(\x)$ is strongly convex and $D-I$ is Lipschitz continuous, where $I$ is the identity operator.
In contrast, we require $D$ to be contractive and we do not require strong convexity of $f$.
In fact, in our case, $f$ is not strongly convex since the sensing matrix $\boldsymbol{\Phi}$ has a non-trivial null space.

Various methods for Gaussian denoising of ECG signals have been explored in the literature; see \cite{chatterjee2020review} for a review.
The state-of-the-art techniques are optimization-based, e.g. TV, multiresolution analysis methods such as wavelets, empirical mode decomposition methods etc.
A combination of these methods is sometimes used \cite{kumar2018denoising}.
Further, nonlocal means (NLM) denoising has also been found to be promising \cite{tracey2012nonlocal}.
However, to the best of our knowledge, there is no work that determines whether any of these denoisers are contractive.
Can we design a high-quality contractive ECG signal denoiser?
In the next section, we show that this can indeed be done.
Specifically, we design a Gaussian denoiser which takes the form $D(\z) = \W \z$, where $\W$ is a symmetric matrix.
The resulting PnP-PGD algorithm outperforms state-of-the-art methods for ECG.
\begin{figure*}[h]
	\centering
	{\includegraphics[width=\linewidth]{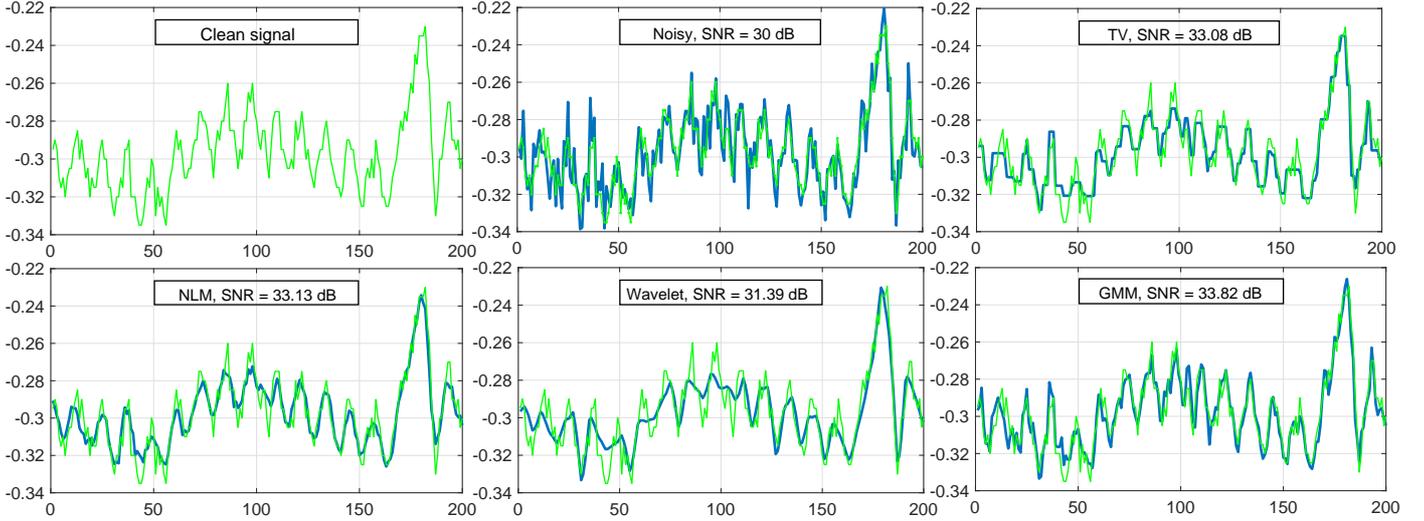}}
	\caption{Comparison of the denoising performance using various state-of-the-art signal denoisers on an ECG signal of length $N = 200$. The clean signal (green) is overlaid in every plot for comparison. Note that the denoised signal using GMM has more structural similarity with the clean signal.}
	\label{denoising_exp1}
\end{figure*}

\section{GMM Denoiser}
\label{sec:denoiser}

Our ECG denoiser is inspired from an observation that was made in the context of images \cite{ZW2011,TBF2018}: A small patch of some fixed size belonging to a clean (i.e. noiseless) image can be well-modeled as a random vector having a Gaussian Mixture Model (GMM) as its density.
Such a density can be learned by fitting a GMM to a large collection of patches extracted from a set of clean images, usually belonging to a common class (e.g. face images).
We apply this idea to model patches belonging to ECG signals.
Specifically, we extract a large collection of patches of length $P \ll N$ from a set of noiseless ECG signals as our training data set.
This is used to fit a GMM density (with a pre-determined number of components $K$) using the expectation-maximization (EM) algorithm.
Essentially, we model clean ECG patches of length $P$ as random vectors $\v \in \mathbb{R}^P$ drawn from this learned GMM density, which we denote by $p(\v)$.
For $j=1,\ldots,K$, let $\alpha_j \geqslant 0$ be the mixture coefficient of the $j$-th component, $\boldsymbol{\mu}_j \in \mathbb{R}^P$ be the mean and $\boldsymbol{\Sigma}_j$ be the positive definite covariance matrix.
Then $p(\v)$ is given by
\begin{equation}
\label{gmm}
p(\v) = \sum_{j=1}^{K} \alpha_j \mathcal{N}(\v;\boldsymbol{\mu}_j,\boldsymbol{\Sigma}_j),
\end{equation}
where $\mathcal{N}$ denotes a Gaussian density function.

How can this model be used for denoising an ECG signal corrupted with Gaussian noise?
Again, we borrow from a patch-based denoising framework that is quite popular in image processing \cite{ZW2011,TBF2018}.
At a high level, this framework consists of the following steps:
\begin{enumerate}[label=\arabic*., labelwidth=!, labelindent=0pt]
\item Extract all possible patches of length $P$ from the noisy signal; if the signal length is $N$, then there are $N$ such patches (we apply circular padding to the signal).
If $\z$ denotes the noisy signal, the collection of patches is given by $\P_1 \z, \ldots, \P_N \z$, where $\P_i : \mathbb{R}^N \to \mathbb{R}^P$ is the linear operator that extracts the patch starting at the $i$-th location.
This is defined as the segment $(z_i, z_{i+1},\ldots,z_{i+P-1})$.
\item Denoise each patch independently by computing a Bayesian estimate of its corresponding clean patch under an additive Gaussian noise model, using $p(\v)$ as the prior distribution of clean patches.
Letting $G$ denote the denoising operator, the collection of denoised patches is given by $G(\P_1 \z),\ldots,G(\P_N \z)$.
\item Place each denoised patch back into its corresponding location in the signal. Each sample location $i \in \{1,\ldots,N\}$ belongs to $P$ overlapping denoised patches; take the average of the $P$ values at this location as the estimate of the $i$-th sample of the denoised signal.
This completes the overall denoising process, which is given by
\begin{equation}
\label{patch-denoiser}
D(\z) = \frac{1}{P} \sum_{i=1}^N \P_i^\top G \big( \P_i \z \big).
\end{equation}
\end{enumerate}
The patch denoiser $G$ forms the core of this framework, and the overall denoising performance depends on the performance of $G$.
One approach to incorporate the Bayesian prior $p(\cdot)$ in the patch denoising is to take $G(\cdot)$ as the maximum a-posteriori (MAP) estimator of the clean patch under a Gaussian noise model.
That is, for a noisy patch $\u = \v + \n$ (where $\n$ is Gaussian noise), we can define $G(\u)$ to be the mode of the conditional density $p(\v|\u)$.
However, it is known that this cannot be computed in closed form when the prior $p(\v)$ is a GMM \cite{ZW2011}.
Instead, motivated by \cite{TBF2018}, we choose $G$ to be the minimum mean-squared-error (MMSE) estimator of the clean patch:
\begin{equation}
\label{mmse-est}
G(\u) = \mathbb{E}[\v | \u].
\end{equation}
The theorem below gives a closed-form expression for $\mathbb{E}[\v | \u]$.

\begin{theorem}[\cite{TBF2018}]
\label{mmse-estimate}
Consider the additive noise model $\u = \v + \n$, where $\n$ is zero-mean Gaussian noise having variance $\sigma^2$.
Suppose $\v$ has the GMM density given by \eqref{gmm}.
Then,
\begin{equation*}
\label{gmm_mmse}
G(\u) = \left( \sum_{j=1}^{K} \beta_j(\u) \C_j \right) \u,
\end{equation*}
where $\C_j = \boldsymbol{\Sigma}_j(\boldsymbol{\Sigma}_j + \sigma^2\I)^{-1}$ and
\begin{equation*}
\beta_j(\u) = \frac{\alpha_j\mathcal{N}(\u;\boldsymbol{\mu}_j,\boldsymbol{\Sigma}_j+\sigma^2\I)}{\sum_{l=1}^{K}\alpha_l\mathcal{N}(\u;\boldsymbol{\mu}_l,\boldsymbol{\Sigma}_l+\sigma^2\I)}.
\label{gmm_beta}
\end{equation*}
\end{theorem}

To summarize, the overall GMM denoiser $D$ is given by
\begin{equation}
\label{D-adaptive}
D(\z) = \frac{1}{P} \sum_{i=1}^N \P_i^\top \left[ \left( \sum_{j=1}^{K} \beta_j \big( \P_i \z \big) \C_j \right) \P_i \z \right].
\end{equation}
In order to gauge the quality of the GMM denoiser, we perform a denoising experiment on a noisy ECG signal.
Specifically, we compare its performance with the following ECG signal denoising schemes: TV \cite{C2013}, NLM \cite{tracey2012nonlocal}, and wavelet-$\ell_1$ regularization.
In Figure.~\ref{denoising_exp1}, we show a segment of signal $\# 115$ (assumed noiseless) from the Physionet MIT-BIH Arrhythmia Database \cite{MIT2000,moody2001impact,LKP2000}.
The segment has length $N = 200$.
We add white Gaussian noise such that the signal-to-noise ratio (SNR) of the noisy signal is $30$ dB.
The SNR for an estimated signal $\hat{\x}$ (here, the denoised signal) with respect to a reference signal $\x$ (here, the clean signal) is defined as
\begin{equation*}
\text{SNR} = \ 10 \log_{10}\Bigg(\frac{||\x||^2}{||\x - \hat{\x}||^2}\Bigg).
\end{equation*}
A higher SNR value indicates a better estimation quality.
The denoised signals obtained using the aforementioned denoising schemes are shown in Figure.~\ref{denoising_exp1}.
Observe that the GMM denoiser yields the highest SNR of all the methods; the visual quality is considerably better compared to TV and wavelet, and comparable to NLM.

For a more extensive comparison, we repeat this experiment for different SNR values of the noisy signal.
The SNR values of the denoised signal are noted in Table \ref{denoising_table}.
Again, it is observed that the GMM denoiser outperforms the other denoisers, while NLM is the second-best method.
In fact, for high noise levels (SNR of $15$ and $20$ dB), the gap in performance between GMM and NLM is quite high.
A possible explanation is that when the noise level is high, reliable computation of weights for NLM is difficult and can result in spikes in the denoised signal.

\begin{table}
\centering
	\caption{Denoising performance (SNR in dB). The signal length is $N = 200$.}
	\scalebox{1.1}{
		\begin{tabularx}{0.75\linewidth}{XXXXX}
			\hline
			Noisy             & TV & NLM & Wavelet &GMM \\
			\hline
			 $15$  & $25.721$ &$26.257$&$23.216$ &$\mathbf{27.492}$  \\
			 	\hline
			$20$     & $26.901$ & $27.023$&$25.861$  & $\mathbf{28.373}$  \\
				\hline
			$25$     & $29.252$ & $29.415$& $29.157$& $\mathbf{29.646}$ \\
				\hline
			$30$    & $33.081$ & $33.139$&  $31.396$& $\mathbf{33.819}$  \\
			\hline
			$35$     & $35.322$ & $35.892$&$32.699$& $\mathbf{36.276}$  \\
				\hline
		    $40$     & $40.391$ &$40.455$& $33.681$& $\mathbf{41.262}$ \\
				\hline
	\end{tabularx}
}
	\label{denoising_table}
\end{table}

\section{Convergence Analysis}
\label{sec:conv}

Recall that we would like $D$ to be contractive; however, due to the complexity of the expression in \eqref{D-adaptive}, it is difficult to determine whether this is the case.
Fortunately, while using it as part of the larger PnP-PGD framework, we can modify the denoiser to make it contractive using a simple trick.
Note that the coefficients $\beta_j(\P_i \z)$ in \eqref{D-adaptive} are nonnegative and sum to $1$ for each $i$.
Consider the situation where we replace the $\beta_j(\P_i \z)$'s by some fixed universal constants $b_{ji}$ that do not depend on $\z$, but have the same properties: $b_{ji} \geqslant 0$ for all $i,j$, and $\sum_j b_{ji} = 1$ for all $i$.
Then $D(\z)$ becomes a linear function of $\z$.
In fact, we can write $D(\z) = \W \z$, where $\W \in \mathbb{R}^{N \times N}$ is given by
\begin{equation}
\label{W}
\W = \frac{1}{P} \sum_{i=1}^N \P_i^\top \left( \sum_{j=1}^{K} b_{ji} \C_j \right) \P_i.
\end{equation}
\begin{theorem}
\label{W-contractive}
Let $\W$ be defined as in \eqref{W}, where $b_{ji} \geqslant 0$ for all $i,j$, and $\sum_{j=1}^K b_{ji} = 1$ for all $i$.
Suppose $N$ is a multiple of $P$.
Then the largest eigenvalue of $\W$, $\lambda_{\mathrm{max}}(\W)$, is $<1$.
Consequently, the denoiser $D(\z) = \W \z$ is contractive, with the constant $\delta$ being $\lambda_{\mathrm{max}}(\W)$.
\end{theorem}

The proof is given in the appendix.
Note that the requirement for $N$ to be a multiple of $P$ is not too restrictive, since we can pad the signal if it is not.
We only need to find suitable coefficients $b_{ij}$ to replace $\beta_j(\P_i \z)$ in \eqref{D-adaptive}.
This can be done as follows.
We first run a small number $T$ (say, $T=10$ or $20$) of PnP-PGD iterations using the coefficients $\beta_j(\P_i \z)$ in the denoiser, to get an intermediate estimate $\x_T$.
We then set $b_{ij} = \beta_j(\P_i \x_T)$ for all $i$ and $j$.
That is, we fix the $b_{ij}$'s as the coefficients obtained from the intermediate point $\x_T$.
The subsequent PnP-PGD iterations are run using the denoiser in \eqref{W} that uses these fixed coefficients.
Since $\W$ is contractive, it follows from Theorem \ref{pnp-ista-conv} that the sequence $\x_{T+1},\x_{T+2},\ldots$ converges to some fixed point $\x^\ast$.
Consequently, the PnP-PGD iterations $\x_1,\x_2,\ldots$ converge to some fixed point $\x^\ast$.

The idea behind fixing the coefficients after $T$ iterations is that as $k$ increases, $\x_k$ is expected to become more refined (in the sense of looking similar to the unknown signal $\x$); therefore, the coefficients $\beta_j(\P_i \x_T)$ after $T$ iterations would be good enough to use for all subsequent iterates as well.
In fact, this scheme has been used in PnP algorithms for image restoration \cite{sreehari2016plug,NGC2021}.

We note that although the patch denoiser $G$ in \eqref{mmse-est} is an MMSE estimator, the overall image denoiser $D$ is not. Therefore existing convergence results for PnP with MMSE denoisers, e.g. \cite{xu2020provable}, do not apply to our case.
We make an important remark on the similarity and differences of the proposed GMM denoiser with the denoiser in \cite{teodoro2017scene,TBF2018}.
Indeed, the idea of our GMM denoiser is inspired by that in \cite{teodoro2017scene,TBF2018}.
However, there are a couple of subtle differences:
\begin{itemize}
\item The denoiser in \cite{teodoro2017scene,TBF2018} is \textit{scene-adapted}, in the sense that the GMM distribution is tailored for the specific scene being reconstructed.
This is possible because the application considered there is hyperspectral image sharpening, in which a complementary image of the same scene is available to obtain training data tailored for that scene.
In contrast, in this work, we learn just one GMM distribution that is kept common for all the signals being reconstructed.
\item The method used to replace $\beta_j(\P_i \z)$ by a fixed coefficient $b_{ij}$ is different in our paper as compared to [2]. This is because the approach in [2] fundamentally relies on the availability of a complementary image, and thus cannot be applied to compressive sensing. In particular, in our paper, we take the $b_{ij}$'s to be the coefficients acquired from a surrogate signal that is obtained by running a few iterations of the PnP-PGD algorithm; this idea was inspired by [Sreehari].
On the other hand, in [2], the $b_{ij}$'s are taken to the coefficients obtained from the complementary image (multispectral or panchromatic image).
\end{itemize}

\begin{figure*}[t!]
	\centering
	{\includegraphics[width=\linewidth]{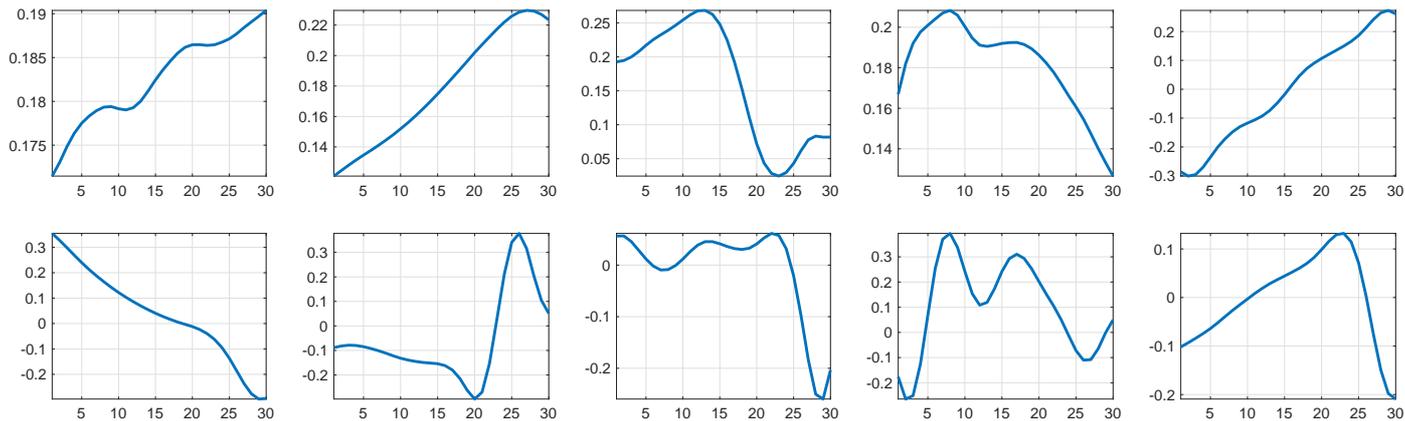}}
	\caption{Ten randomly selected eigenvectors corresponding to the largest few eigenvalues of the covariance matrices from the learned GMM model.}
	\label{cov_low}
\end{figure*}

\begin{figure*}[t!]
	\centering
	{\includegraphics[width=\linewidth]{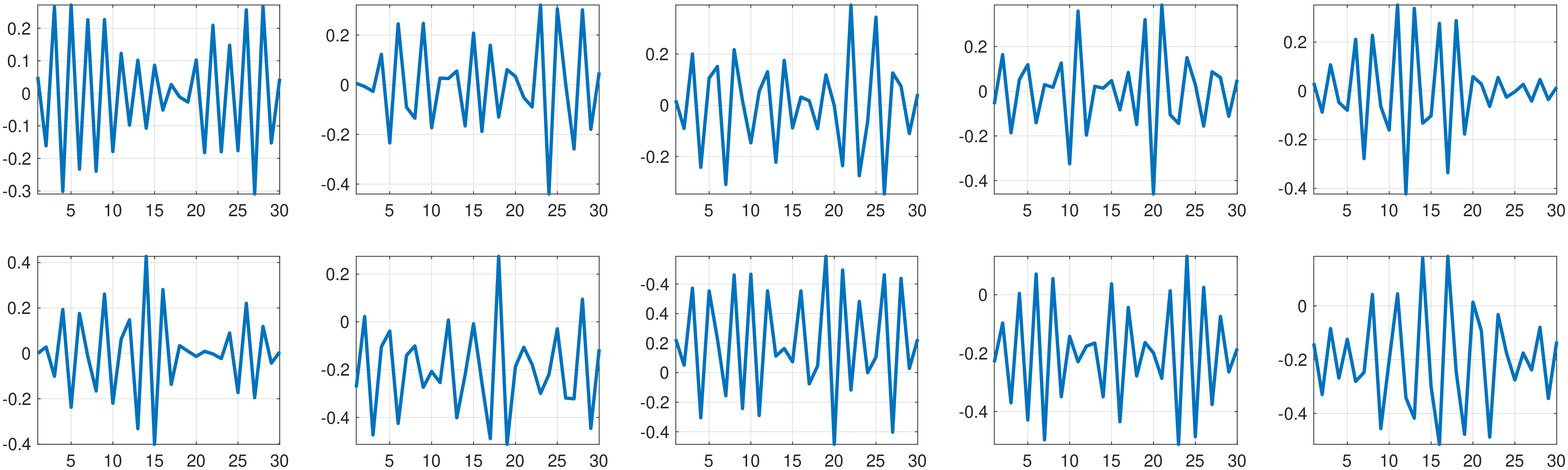}}
	\caption{Ten randomly selected eigenvectors corresponding to the smallest few eigenvalues of the covariance matrices from the learned GMM model.}
	\label{cov_high}
\end{figure*}

\begin{figure*}[t!]
	\centering
	{\includegraphics[width=0.95\linewidth]{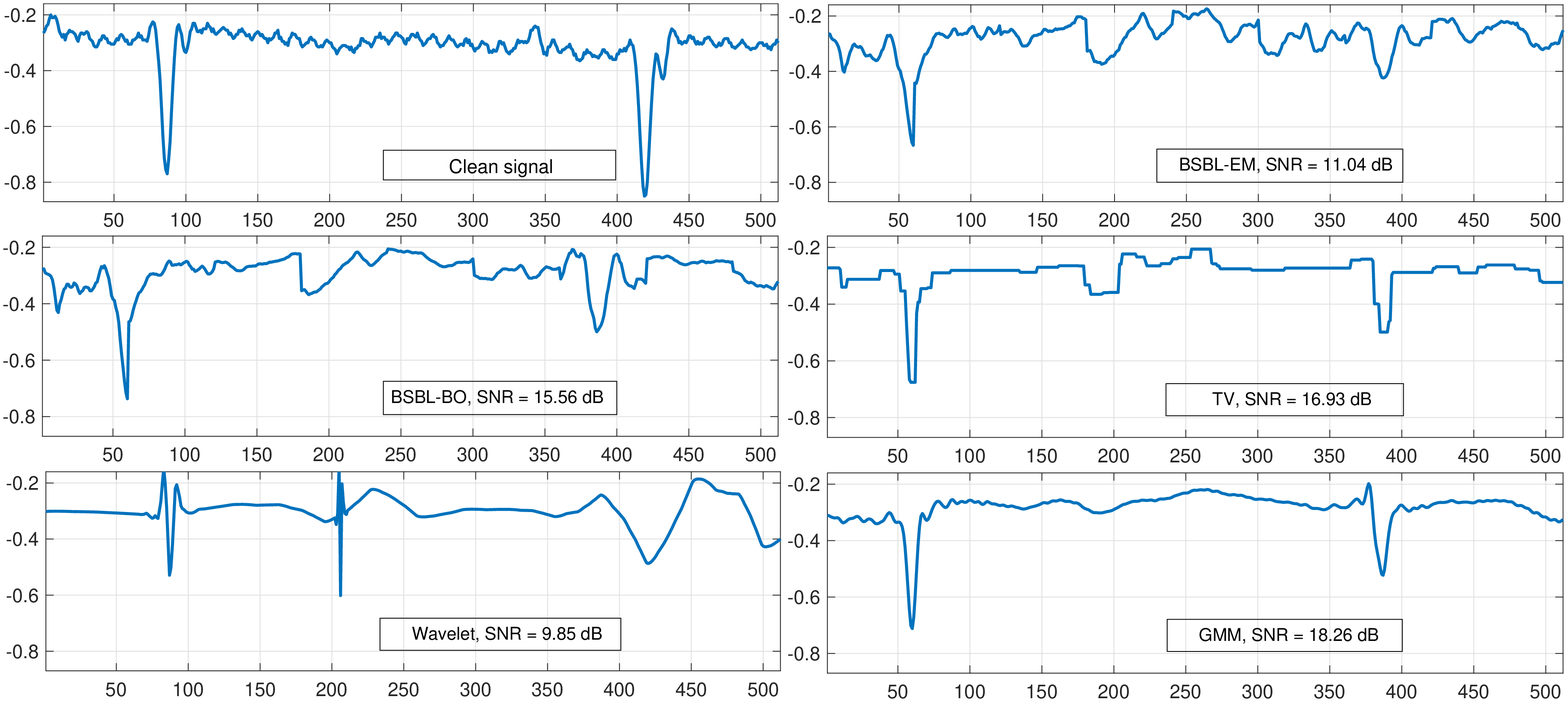}} 
	\caption{Visual comparison of a CS recovered ECG signal with $10\%$ measurements and no additive noise. The length of the original signal is $N=512$. The proposed method produces more structurally similar (with the clean signal) output even under lower number of observations.}
	\label{cs_reconstruction_10}
\end{figure*}

\begin{figure*}[t!]
	\centering
	{\includegraphics[width=\linewidth]{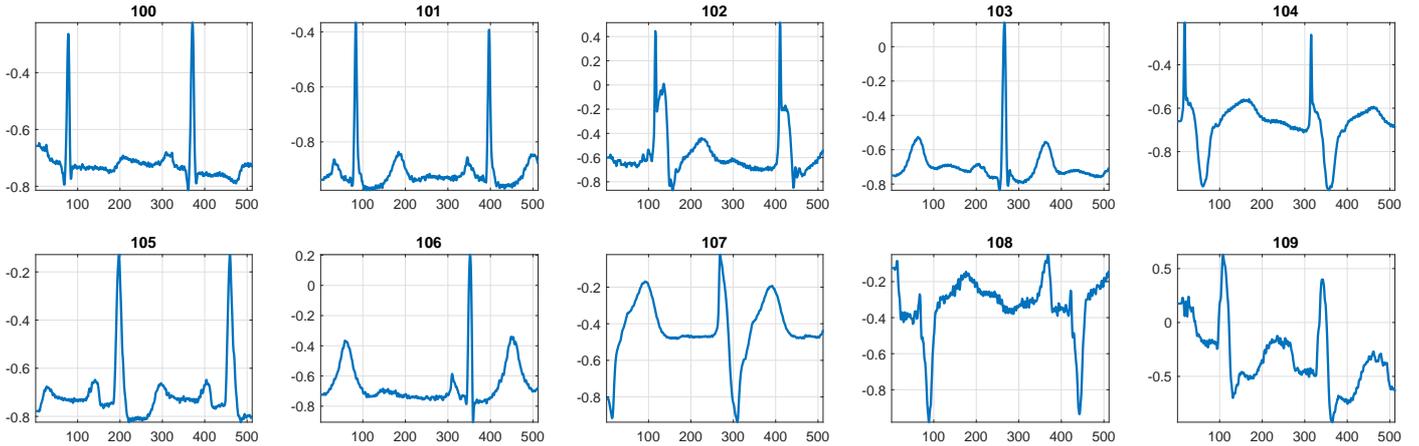}}
	\caption{Different test signals from the MIT-BIH Arrhythmia database.}
	\label{test_signals}
\end{figure*}

\section{Experimental Results}
\label{sec:experiments}

\underline{\textbf{Database:}} To validate the proposed PnP-PGD method for ECG CS recovery, we use a subset of the data from the Physionet MIT-BIH Arrhythmia Database \cite{MIT2000,moody2001impact,LKP2000}.
Every file in the database consists of two lead recordings sampled at 360Hz with 11 bits per sample of resolution. It contains 48 half-hour excerpts of two-channel ambulatory ECG recordings, obtained from 47 subjects studied by the BIH Arrhythmia Laboratory.
 
\underline{\textbf{Metrics:}} We quantify the performance of the proposal using following metrics: SNR, which is defined in Section \ref{sec:denoiser}), and mean-squared error (MSE), which is defined below.
\begin{equation*}
\begin{aligned}
\text{MSE} = & \ \frac{1}{N} ||\x - \hat{\x}||^2,
\end{aligned}
\end{equation*}
where $\hat{\x}$ and $\x$ are the reconstructed and original ECG signals, respectively. Note that here we are assuming the Physionet signals are the true signals $\x$; in reality these signals also contain noise, which the metrics above neglect, though at most SNRs the (simulated) additive noise dominates \cite{tracey2012nonlocal}.

\underline{\textbf{Compared methods:}} We compare with the following state-of-the-art methods: BSBL-BO \cite{zhang2011sparse}, BSBL-EM \cite{zhang2013extension}, sparse prior on the ECG wavelet representation \cite{polania2014weighted}, and TV regularization \cite{liu2013multi}.
We tuned the parameters of all the methods so that maximum SNR is obtained.
The codes are used from the publicly available sites \cite{zhang2012compressed,zhang2011sparse,zhang2013extension}.
All simulations were performed using MATLAB (R2021a) on a Quad core, 3.80GHz machine with 32GB RAM.

The sensing matrix $\boldsymbol{\Phi}$ is constructed by randomly drawing each entry from the standard normal distribution $\mathcal{N}(0,1)$, and applying an orthonormalization step to ensure that the rows of $\boldsymbol{\Phi}$ are orthonormal \cite{pant2014new}.
We trained the GMM on the set of all possible overlapping patches of size $P=30$ extracted from signal \#$104$ in the dataset \cite{MIT2000}, which is of length $10,800$.
The number of GMM components $K$ is set to be $10$. The training time is found to be $2.86$s.
In all the experiments on CS reconstruction, we terminate the PnP-PGD algorithm after $150$ iterations since we observed that the algorithm stabilizes by then.

In addition to the denoising results reported in Section \ref{sec:denoiser}, the results in the subsequent sections support the claim that GMM is a good prior for modeling ECG signal patches. Note that the entire training process can be done offline.

\subsection{Goodness of GMM Modeling}
In this experiment, we evaluate the goodness of GMM modeling by visualizing some of the eigenvectors of the covariance matrices from the learned GMM distribution.
This type of visualization is commonly utilized in papers on image processing, e.g. \cite{ZW2011}.
In particular, in \cite{ZW2011} it is observed that the eigenvectors corresponding to the largest few eigenvalues (of each covariance matrix) are relatively smooth and capture the large-scale structure of the patches, whereas the eigenvalues corresponding to the smallest few eigenvectors contain many fluctuations and thus capture the local structure.
If the GMM is a good model, we expect to see a similar trend for ECG signal patches.

In our case, we fit a GMM with $K=10$ components on ECG patches of size $30$.
In Figure \ref{cov_high}, we plot $10$ randomly selected eigenvectors (having unit norm) corresponding to eigenvalue indices $\geqslant 28$ (i.e. largest few eigenvalues) from the fitted GMM model.
In Figure \ref{cov_high}, we show similar plots but for eigenvalue indices $\leqslant 3$ (i.e. smallest few eigenvalues).
It is evident that the expected trend described in the previous paragraph holds true in practice: the richness of textures, fluctuations and other local structures is captured by the signals in Figure \ref{cov_high}, while most of the large-scale details are captured by the signals in Figure \ref{cov_low}.

\subsection{Recovery from Noiseless Measurements}
We study the signal reconstruction performance of our method (under zero noise), especially when the number of measurements $M$ is much lesser than $N$. In Figure \ref{cs_reconstruction_10}, we show a segment of the original ECG signal $\#100$  from \cite{moody2001impact}  and its reconstruction obtained using our proposed method with $10\%$ measurements.  We can see that the recovered signal using GMM as the denoiser produces visually similar result on comparison with the other methods and also has better performance in terms of SNR.

\subsection{Study of SNR for Different $M$}
\begin{figure}[t!]
	\centering
	{\includegraphics[width=0.9\linewidth]{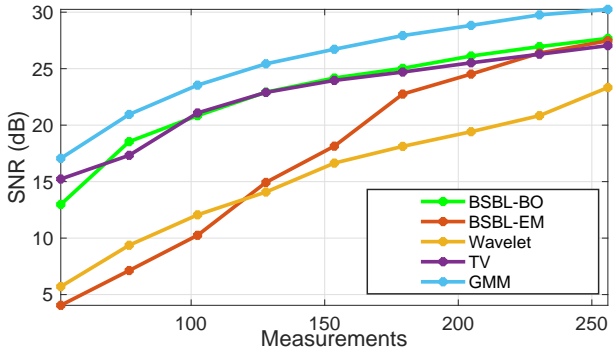}} 
	\caption{Average SNR vs. measurements ($M$) over $500$ runs, with no additive noise. The length of the original ECG signal is $N=512$.}
	\label{SNRvsM}
\end{figure}

We next perform an exhaustive experiment where we vary the number of measurements ($M$) while fixing the length of the signal. The signal  $\# 105$ of length $N = 512$ is used for the experiment. The results are reported in Figure~\ref{SNRvsM}. Each instance of the SNR values shown is obtained by averaging of $500$ independent trials. We note that our proposed method achieves the best recovery among all the methods.
\subsection{Study of Average SNR over Different Signals}

\begin{figure}[t!]
	\centering
	{\includegraphics[width=0.9\linewidth]{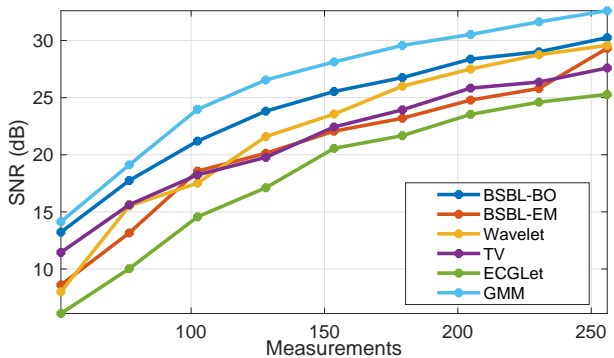}}
	\caption{Average SNR vs. measurements ($M$) averaged over the $10$ ECG signals in Figure \ref{test_signals}. The measurements do not contain any additive noise.}
	\label{snrvsm}
\end{figure}

In this section, we study the performace of the proposed method with that of various ECG signals from MIT-BIH database \cite{moody2001impact}.
We consider a larger dataset of $10$ signals (\#100 to \#109), shown in Figure \ref{test_signals} for performance evaluation.
All the signals are of length $N = 512$.
We also compare with a more recent method, ECGLet \cite{ansari2019wnc}, in this section.
We measure the average SNR (over all the $10$ test signals) of the reconstructed signal for different values of $M$.
The results are reported in Figure \ref{snrvsm}.
Note that the proposed method yields the highest SNR among all the methods considered, for every $M$.

\subsection{Effect of Compression Ratio}
\begin{figure}[t!]
	\centering
	{\includegraphics[width=0.9\linewidth]{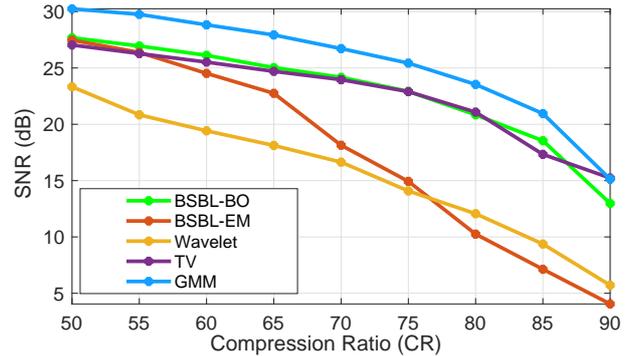}} 
	\caption{Average SNR vs. CR over $500$ runs. The length of the original ECG signal is $N=512$.}
	\label{SNRvsCR}
\end{figure}

In this section we investigate the effect of compression ratio (CR) on the quality of the reconstructed ECG signals. The CR is defined as:
\begin{equation}
	\text{CR} \ = \frac{N-M}{N} \times 100
\end{equation}
where $N$ is the length of the original signal and $M$ is the length of the compressed signal.  
For each value of CR, we repeated the experiment $500$ times, and in each time, the
sensing matrix was randomly generated \cite{zhang2012compressed}. Figure~\ref{SNRvsCR}, shows the variation of SNR with CR. It is worth noting that we obtain superior performance over the whole range of CR values. The input signal is a segment of ECG signal $\#105$ from the  MIT-BIH database \cite{moody2001impact}.

\begin{figure}[t!]
	\centering
	{\includegraphics[width=0.9\linewidth]{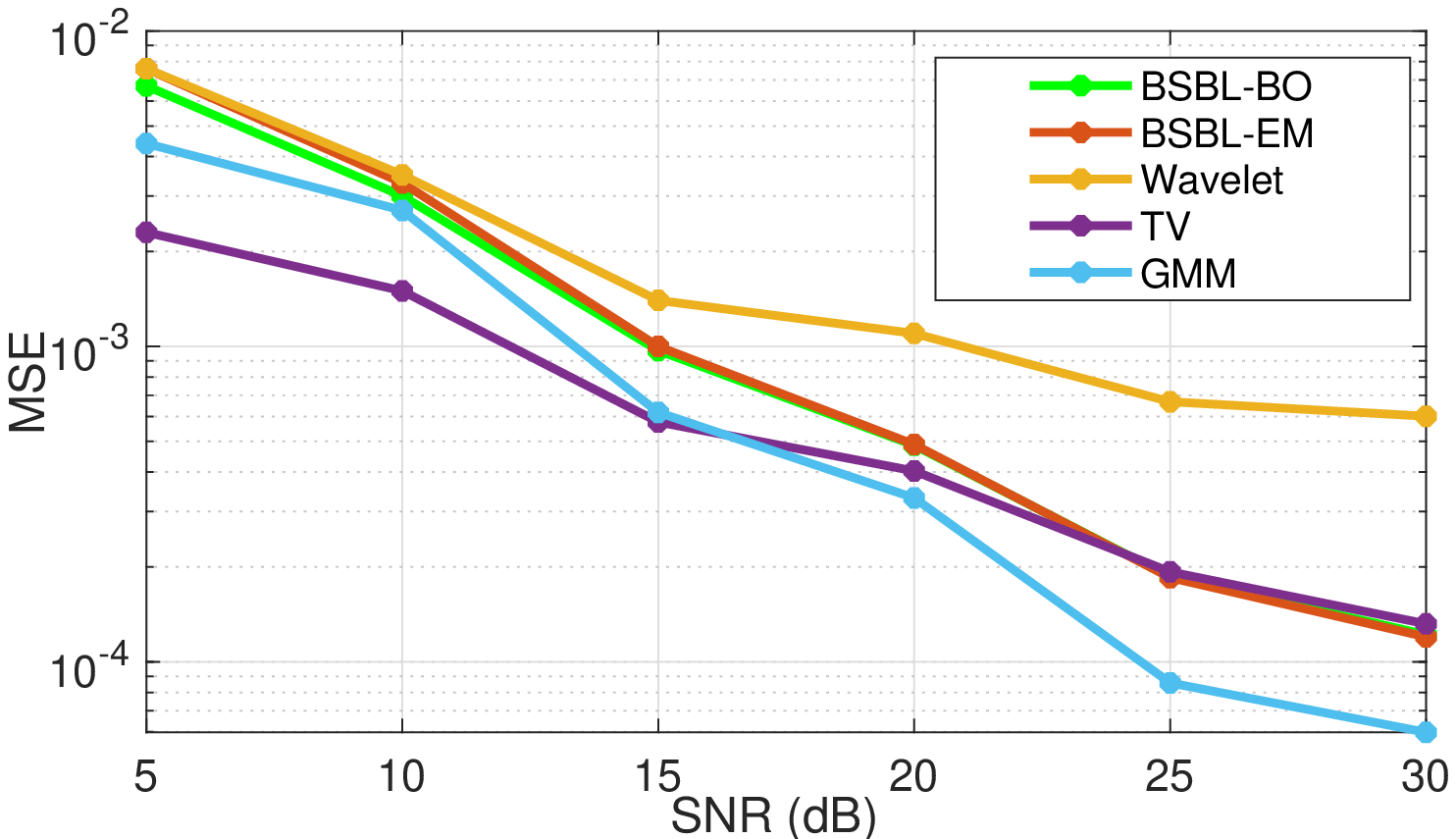}} 
	\caption{Average SNR vs. MSE over $500$ runs. The length of the original ECG signal is $N=512$.}
	\label{fig:SNR-plot}
\end{figure}

\begin{figure*}[t!]
	\centering
	{\includegraphics[width=0.95\linewidth]{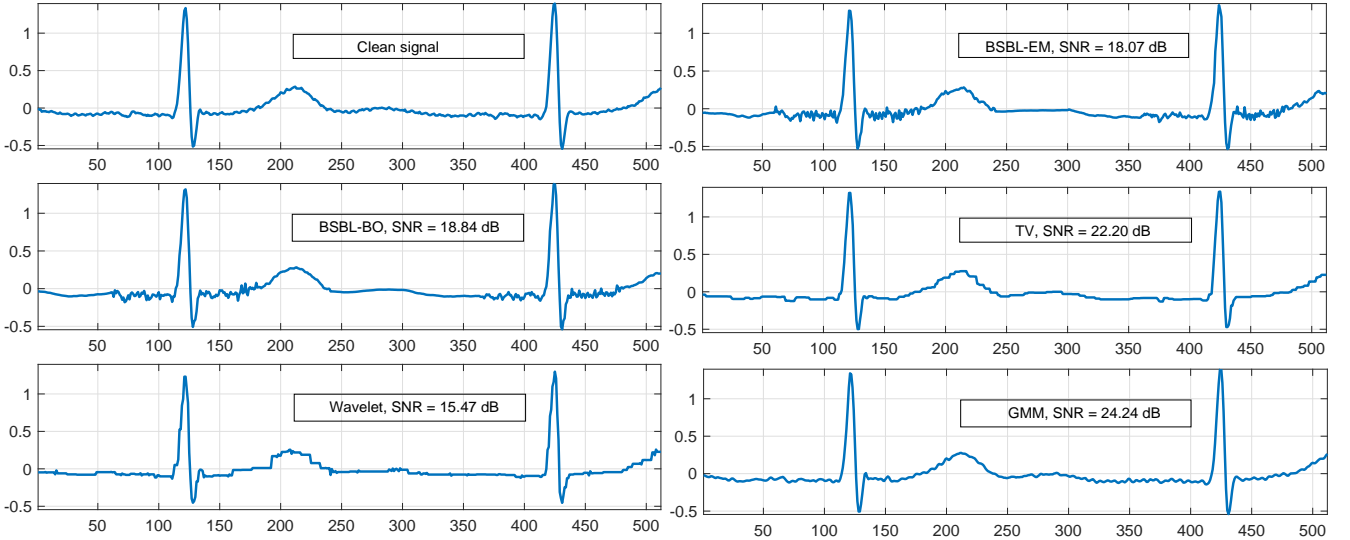}}
	\caption{Visual comparison of a CS recovered ECG signal with $50\%$ measurements and additive Gaussian noise. The length of the original signal is $N=512$. The proposed method is able to capture the minute variations in the signal faithfully.}
	\label{SNR_noisy}
\end{figure*}

\subsection{Recovery from Noisy Measurements}

From previous experiments we notice that the proposed method performs well when compared with the other methods in noiseless scenario, i.e. when $\n=\mathbf{0}$ in \eqref{eq:model}. Now we examine the recovery performance of our method from noisy compressed measurements. 
The signal $\# 103$ of length $N = 512$ and $M= 256$ is used for this experiment.
In Figure \ref{fig:SNR-plot}, we plot the MSE of the recovered signal as a function of the noise level in the input $\y$ (specified in terms of the SNR of the input).
The values reported are averaged over $500$ independent trials.
To simulate the noisy measurements, we followed the approach in \cite{tracey2012nonlocal}. It is evident that the proposed method produces quality reconstructions under noisy measurements.
Finally we show a reconstruction result from noisy measurements in Figure.~\ref{SNR_noisy}. In the experiment, we add random Gaussian noise to the compressed measurements. On comparison, none of the three methods in Figure \ref{SNR_noisy} are able to completely mitigate the effect of noise. However, our method performs the best by a significant margin, resulting in an SNR of $24.24$ dB in the recovered signal, and is visually similar to the original signal.
We observed that in low SNR scenarios, TV acts as a better denoiser than GMM, which might explain why the CS reconstruction performance is higher for TV as compared to the proposed method when the input SNR is low.


\subsection{Numerical Convergence}

\begin{figure*}
	\centering
	{\includegraphics[width=0.97\linewidth]{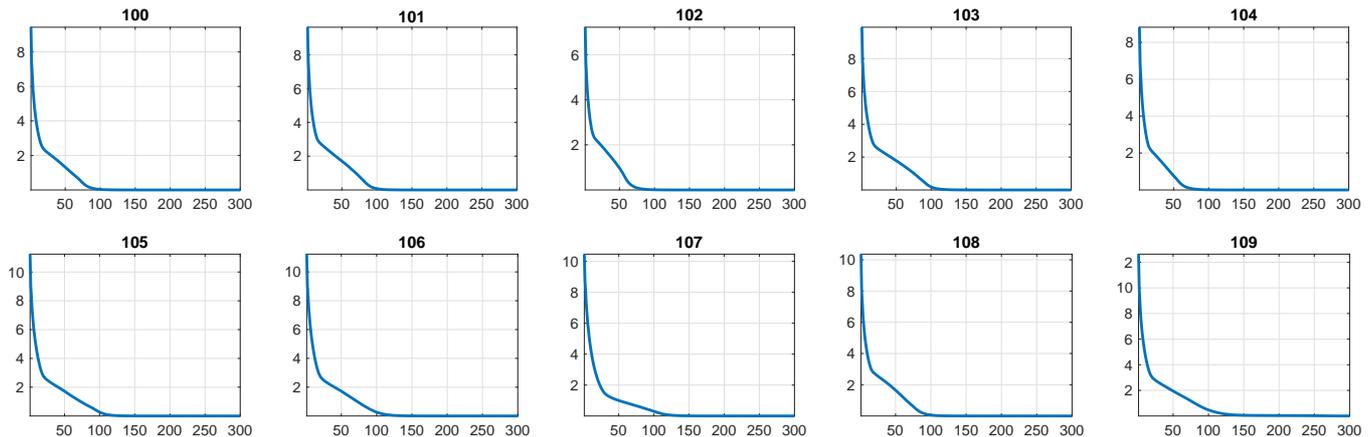}}
	\caption{Plots of $\norm{\x_{k+1} - \x_k}$ vs $k$ in the proposed algorithm (with $M = 0.5 N$) for the test signals in Figure \ref{test_signals}.}
	\label{iter_conv}
\end{figure*}

In this section, we numerically verify the convergence of the proposed PnP-PGD algorithm.
For all the signals in Figure \ref{test_signals}, we use PnP-PGD for reconstruction from $M=0.5N$ measurements.
As explained in Section \ref{sec:conv}, we take the surrogate signal as the signal obtained after running the algorithm for $T = 10$ iterations.
The plots of $\norm{\x_{k+1} - \x_k}$ as a function of $k$ are shown in Figure \ref{iter_conv} for each of the signals.
Note that $\norm{\x_{k+1} - \x_k}$ decays to $0$ as $k$ increases, which is a necessary condition for convergence.

\begin{table*}[t!]
\centering
	\caption{Performance comparison with SDAE on FECG data, collected from Physionet database (patient id: ecgca154) by varying the compression ratio.}
	\scalebox{1.1}{
		\begin{tabularx}{0.78\linewidth}{XXXXXXXX}
			\hline
			CR             & 25 & 37.5 & 50 &62.5&75&87.5&93.75 \\
			\hline
			 SDAE  & $37.69$ & $36.73$ & $35.39$& $32.40$ &$26.78$& $\textbf{23.56}$& $\textbf{22.29}$ \\
			 	\hline
			GMM    & $\textbf{42.85}$ & $\textbf{39.56}$&$\textbf{37.18}$  & $\textbf{33.29}$ &$\textbf{27.16}$ &$20.57$ &$11.44$ \\
				\hline
	\end{tabularx}
}
	\label{deepvsgmm}
\end{table*}
\begin{figure}[t!]
	\centering
	{\includegraphics[width=0.95\linewidth]{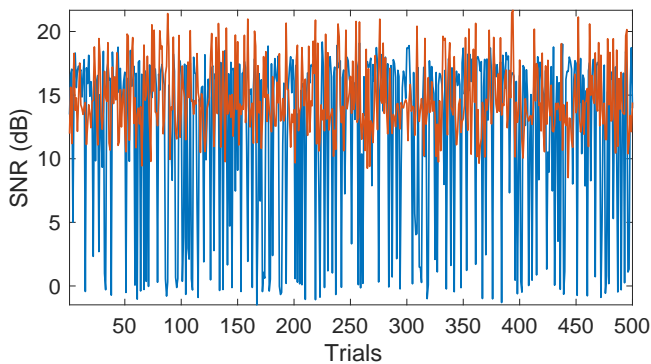}} 
	\caption{Variation of SNR with trials for BSBL-BO (blue) and the proposed method (red). The length of the original ECG signal $N=512$. The proposed method exhibits stable recovery of the ECG signal even for a low compression ratio.}
	\label{stability}
\end{figure}
\subsection{Comparison with Deep Learning}

For completeness, we compare the proposed method with a deep learning method for CS reconstruction, known as SDAE \cite{muduli2016deep}. 
In this experiment, we use the non-invasive Fetal ECG dataset from Physionet, which was used in \cite{muduli2016deep}.
This database contains a series of 55 multichannel abdominal non-invasive fetal electrocardiogram (FECG) recordings, taken from a single subject between 21 to 40 weeks of pregnancy.
We conducted an experiment on the signal with patient id ecgca154.
Table \ref{deepvsgmm} shows the variation of SNR with CR for the proposed method and SDAE.
We note that the proposed method is able to outperform SDAE in all cases except when the CR is very high.

\subsection{Stable Recovery}
We show that the proposed method produces very stable reconstructions. For this experiment, we considered the signal $\#105$ with $10\%$ measurements ($N=512$). We ran $500$ trials of the method, so that its stability can be observed for different realizations of the sensing matrix $\boldsymbol{\Phi}$. In Figure \ref{stability}, we show the SNR variation for  BSBL-BO (blue) and the proposed method (red). We noted a standard deviation of $6.90$ dB in SNR for BSBL-BO and $2.70$ dB for the proposed method.
Thus, the proposed method is more stable as compared to BSBL-BO.
In fact, we observed that the contrast in stability is more pronounced for smaller $M$.

\section{Conclusion}
\label{sec:conclusion}

We introduced a novel framework for recovering ECG signals from compressively sensed measurements.
Our method is based on the plug-and-play (PnP) paradigm that has recently become popular for image restoration problems.
Essentially, the recovery method consists of repeating two main steps -- inverting the forward model, and denoising -- until stability is attained.
We designed a high-quality ECG signal denoiser to be used in the denoising step.
Moreover, we proved that the recovery algorithm is guaranteed to converge.
Importantly, we showed via numerical experiments that our proposed method is superior to current state-of-the-art methods used for ECG CS recovery.

\section{Appendix}
\subsection{Proof of Theorem \ref{pnp-ista-conv}}
Since $\nabla f(\x) = \boldsymbol{\Phi}^\top ( \boldsymbol{\Phi} \x - \y)$, we can write the PnP-PGD algorithm as $\x_{k+1} = S(\x_k)$, where
\begin{equation*}
S(\x) = D \big( \x - \gamma \boldsymbol{\Phi}^\top \boldsymbol{\Phi} \x + \gamma \boldsymbol{\Phi}^\top \y \big).
\end{equation*}
It is enough to prove that the function $S(\cdot)$ is contractive, since the convergence of $(\x_k)$ to some unique fixed point $\x^\ast$ at a linear rate would then follow by the Banach Fixed Point Theorem \cite[Th. 9.23]{W1976}.

Let $\delta < 1$ be the constant in Definition \ref{contractive-map}.
Then for any $\z_1,\z_2 \in \mathbb{R}^N$, we have,
\begin{align*}
\left \lVert S(\z_1) - S(\z_2) \right \rVert &\leqslant \delta \lVert (\z_1 - \gamma \boldsymbol{\Phi}^\top \boldsymbol{\Phi} \z_1) - (\z_2 - \gamma \boldsymbol{\Phi}^\top \boldsymbol{\Phi} \z_2) \rVert \\
&= \delta \lVert (\I - \gamma \boldsymbol{\Phi}^\top \boldsymbol{\Phi}) (\z_1 - \z_2) \rVert \\
&\leqslant \delta \cdot \sigma_{\mathrm{max}}(\I - \gamma \boldsymbol{\Phi}^\top \boldsymbol{\Phi}) \cdot \lVert \z_1 - \z_2 \rVert.
\end{align*}
Let $L = \sigma_{\mathrm{max}}(\boldsymbol{\Phi}^\top \boldsymbol{\Phi})$.
Since $\boldsymbol{\Phi}^\top \boldsymbol{\Phi}$ is positive semidefinite, its singular values are also its eigenvalues.
In particular, its eigenvalues lie in $[0,L]$.
Since $0 < \gamma \leqslant 2/L$, the eigenvalues of $\I - \gamma \boldsymbol{\Phi}^\top \boldsymbol{\Phi}$ lie in $[-1,1]$.
Therefore, $\sigma_{\mathrm{max}}(\I - \gamma \boldsymbol{\Phi}^\top \boldsymbol{\Phi}) \leqslant 1$.
Thus, for all $\z_1,\z_2 \in \mathbb{R}^N$ we have,
\begin{equation*}
\left \lVert S(\z_1) - S(\z_2) \right \rVert \leqslant \delta \lVert \z_1 - \z_2 \rVert.
\end{equation*}
Since $\delta < 1$, the function $S(\cdot)$ is contractive.

\subsection{Proof of Theorem \ref{W-contractive}}
A proof can be found in \cite[Appendix B]{TBF2018}; for completeness, here we give a different and more concise proof.
Note that each $\C_j$ is symmetric positive semidefinite (p.s.d.); hence, for each $i$, the matrix $\mathbf{\B}_i := \sum_j b_{ji} \C_j$ is p.s.d. (as a convex combination of p.s.d. matrices).
By the same logic, we get that $\W$ is p.s.d.
Thus, to show $\lambda_{\mathrm{max}}(\W) < 1$, we only need to show that $\z^\top \W \z < \lVert \z \rVert^2$ for all $\z \in \mathbb{R}^N$.

To prove this, first note that $\lambda_{\mathrm{max}}(\C_j) < 1$ for all $j$.
Since each $\B_i$ is a convex combination of all $\C_j$'s and since $\lambda_{\mathrm{max}}(\cdot)$ is a convex function on the set of symmetric matrices, we have $\lambda_{\mathrm{max}}(\B_i) < 1$ for all $i=1,\ldots,N$.
Let
\begin{equation*}
\delta = \max \big( \lambda_{\mathrm{max}}(\B_1),\ldots,\lambda_{\mathrm{max}}(\B_N) \big).
\end{equation*}
Clearly, $\delta < 1$.
Note that for each $i$, we have $\u^\top \B_i \u \leqslant \lambda_{\mathrm{max}}(\B_i) \lVert \u \rVert^2 \leqslant \delta \lVert \u \rVert^2$.
Therefore, for any $\z \in \mathbb{R}^N$,
\begin{equation*}
\z^\top \W \z = \frac{1}{P} \sum_{i=1}^N (\P_i \z)^\top \B_i (\P_i \z) \leqslant \frac{\delta}{P} \sum_{i=1}^N \lVert \P_i \z \rVert^2,
\end{equation*}
Since $N$ is a multiple of $P$, $\sum_{i=1}^N \lVert \P_i \z \rVert^2$, which is the sum of all patches of length $P$ extracted from $\z$ (using circular padding), is simply equal to $P \lVert \z \rVert^2$.
Thus, we get that $\z^\top \W \z \leqslant \delta \lVert \z \rVert^2 < \lVert \z \rVert^2$ for all $\z \in \mathbb{R}^N$.

\bibliographystyle{elsarticle-num}
\bibliography{citations}
\end{document}